\documentclass[a4paper,fleqn,usenatbib]{mnras}

\usepackage{newtxtext,newtxmath}

\usepackage[T1]{fontenc}
\usepackage{ae,aecompl}

\usepackage{graphicx}	
\usepackage{amsmath}	
\usepackage{amssymb}	
\usepackage{enumitem}
\usepackage{multicol}

\title[Weighing in on the masses of retired A stars]{
Weighing in on the masses of retired A stars with asteroseismology\\
\vspace{0.1in} 
\LARGE K2 observations of the exoplanet-host star HD~212771\thanks{Based on observations collected at La Silla Observatory (ESO, Chile) with the FEROS spectrograph at the 2.2-m telescope under programme 088.C-0892(A).}
}

\author[T.~L.~Campante et al.]
{Tiago L.~Campante,$^{1,2,3}$\thanks{E-mail: campante@bison.ph.bham.ac.uk (TLC)}
Dimitri Veras,$^{4}$
Thomas S.~H.~North,$^{1,2}$
Andrea Miglio,$^{1,2}$
\newauthor
Thierry Morel,$^{5}$
John A.~Johnson,$^{6}$
William J.~Chaplin,$^{1,2}$
Guy R.~Davies,$^{1,2}$
\newauthor
Daniel Huber,$^{7,2,8}$
James S.~Kuszlewicz,$^{1,2}$
Mikkel N.~Lund,$^{1,2}$
Benjamin F.~Cooke,$^{1}$
\newauthor
Yvonne P.~Elsworth,$^{1,2}$
Tha\'ise S.~Rodrigues$^{9,10}$
and Andrew~Vanderburg$^{6}$
\\
$^{1}$School of Physics and Astronomy, University of Birmingham, Edgbaston, Birmingham B15 2TT, UK\\
$^{2}$Stellar Astrophysics Centre (SAC), Department of Physics and Astronomy, Aarhus University, Ny Munkegade 120,\\ DK-8000 Aarhus C, Denmark\\
$^{3}$Institut f\"{u}r Astrophysik, Georg-August-Universit\"{a}t G\"{o}ttingen, Friedrich-Hund-Platz 1, 37077 G\"{o}ttingen, Germany\\
$^{4}$Department of Physics, University of Warwick, Coventry CV4 7AL, UK\\
$^{5}$Space sciences, Technologies and Astrophysics Research (STAR) Institute, Universit\'e de Li\`ege, Quartier Agora,\\ All\'ee du 6 Ao\^ut 19c, B\^at.~B5C, B4000-Li\`ege, Belgium\\
$^{6}$Harvard-Smithsonian Center for Astrophysics, 60 Garden Street, Cambridge, MA 02138, USA\\
$^{7}$Sydney Institute for Astronomy, School of Physics, University of Sydney, Sydney, Australia\\
$^{8}$Institute for Astronomy, University of Hawaii, 2680 Woodlawn Drive, Honolulu, HI 96822, USA\\
$^{9}$Osservatorio Astronomico di Padova -- INAF, Vicolo dell'Osservatorio 5, I-35122 Padova, Italy\\
$^{10}$Dipartimento di Fisica e Astronomia, Universit\`a di Padova, Vicolo dell'Osservatorio 2, I-35122 Padova, Italy
}

\date{Accepted XXX. Received YYY; in original form ZZZ}

\pubyear{2016}

\begin{document}
\label{firstpage}
\pagerange{\pageref{firstpage}--\pageref{lastpage}}
\maketitle

\begin{abstract}
Doppler-based planet surveys point to an increasing occurrence rate of giant planets with stellar mass. Such surveys rely on evolved stars for a sample of intermediate-mass stars (so-called retired A stars), which are more amenable to Doppler observations than their main-sequence progenitors. However, it has been hypothesised that the masses of subgiant and low-luminosity red-giant stars targeted by these surveys --- typically derived from a combination of spectroscopy  and isochrone fitting --- may be systematically overestimated. Here, we test this hypothesis for the particular case of the exoplanet-host star HD~212771 using K2 asteroseismology. The benchmark asteroseismic mass ($1.45^{+0.10}_{-0.09}\:\text{M}_{\sun}$) is significantly higher than the value reported in the discovery paper ($1.15\pm0.08\:\text{M}_{\sun}$), which has been used to inform the stellar mass-planet occurrence relation. This result, therefore, does not lend support to the above hypothesis. Implications for the fates of planetary systems are sensitively dependent on stellar mass. Based on the derived asteroseismic mass, we predict the post-main-sequence evolution of the Jovian planet orbiting HD~212771 under the effects of tidal forces and stellar mass loss.
\end{abstract}




\section{Introduction}
Studies based on Doppler surveys have suggested an increasing occurrence rate of giant planets with stellar mass \citep[][]{JohnsonMdwarf,LovisMayor,Bowler10,Johnson10}, taken as supporting evidence of the core-accretion model of planet formation. Such studies rely, at the higher-mass end (i.e., $M\ga1.5\,\text{M}_{\sun}$), on the evolved counterparts of A- and early F-type stars \citep[e.g.,][]{Hatzes03,Setiawan05,Reffert06,Johnson07,Sato08,Wittenmyer11,TAPAS}. These stars, collectively known in the literature as \textit{retired A stars}, exhibit significantly slower surface rotation rates than their main-sequence progenitors \citep[e.g.,][]{doNascimento12,Garcia14}, hence becoming more amenable to Doppler-based planet surveys. The validity of the above result is, nevertheless, subject to our ability to measure accurate masses for evolved stars.

The masses of subgiant and low-luminosity red-giant stars targeted by Doppler surveys discussed in the literature are estimated by fitting the outputs of stellar evolutionary models to a set of observables, typically the luminosity, and the spectroscopically-determined effective temperature and metallicity. Recently, though, \citet{Lloyd11} has called into question these mass estimates, arguing that the selection criteria used to define samples of evolved stars for Doppler-based planet surveys should instead have led to a sample dominated by lower-mass stars, more likely to have originated from a population of late F/early G dwarfs with masses in the range $1.0$--$1.2\,\text{M}_{\sun}$. The debate over this claim is ongoing \citep{Johnson13,Lloyd13,Schlaufman13,Johnson14,GhezziJohnson}. A resolution to this issue would carry important implications for the way in which masses are estimated for stars in the subgiant and giant branches, especially in the absence of asteroseismic information. Furthermore, it has a direct impact on our understanding of planet occurrence as a function of stellar mass. In that regard, knowledge of giant-planet occurrence rates around intermediate- and high-mass stars is central to accurately predicting the yield of planet imaging surveys \citep[e.g.,][]{CreppJohnson}.

The advent of space-based asteroseismology has vastly benefitted the study of solar-type (i.e., low-mass, main-sequence stars and cool subgiants) and red-giant stars \citep[for a review, see][]{ChaplinMiglio}. These stars exhibit solar-like oscillations, which are stochastically excited and intrinsically damped by turbulence in the outermost layers of a star's convective envelope. Analysis of solar-like oscillations has allowed the precise determination of fundamental stellar properties (e.g., mass, radius and age) for large numbers of field stars, including over a hundred exoplanet hosts \citep[e.g.,][]{Huber13,Campante15,VSA15,CampanteObliquities,Davies16,Lundkvist}. Tests of the accuracy of asteroseismic masses have so far mostly relied on studies of red-giant members of open clusters. Current studies suggest that, especially when stellar-model-based corrections to the large frequency separation ($\Delta\nu$; see Sect.~\ref{sec:asteroseismology} for a definition) scaling relation are applied, asteroseismic masses are compatible with independent mass estimates, showing no evidence of systematic offsets \citep[e.g.,][and references therein]{Miglio16,Sharma16,Stello16}.

Asteroseismology with the repurposed NASA K2 mission \citep[e.g.,][]{K2astero1,K2astero2,Lund16,LundHyades,Miglio16} --- successively targeting different fields along the ecliptic plane --- therefore has the potential to shed new light on the retired A star controversy by providing accurate and precise masses for a number of bright, subgiant and low-luminosity red-giant host stars previously targeted by Doppler surveys. Herein, we address this issue by presenting the first\footnote{\citet{Grunblatt} have recently reported the first newly discovered K2 planet around an asteroseismic host star.} asteroseismic characterisation of a known exoplanet-host star using K2 photometry, deferring an ensemble study to a future publication \citetext{T.~S.~H.~North et al., in prep.}. HD~212771 (EPIC~205924248, HIP~110813, 2MASS~J22270308-1715492) is a bright \citep[$V\!=\!7.60$;][]{Tycho2} subgiant of spectral type G8$\,$IV \citep[][]{spectype}, being amongst the targets of the Doppler-based planet survey of \citet{Johnson07}. It hosts a Jovian planet (with minimum mass $M_\text{p}\sin i\!=\!2.3\pm0.4\,M_\text{Jup}$) in a 373.3-day orbit \citep{discovery}. The stellar mass reported in the discovery paper ($1.15\pm0.08\,\text{M}_{\sun}$) does not place HD~212771 in the retired A star category. However, other recent spectroscopic studies have provided mass estimates for this star in the range $1.22$--$1.60\,\text{M}_{\sun}$ \citep{Mortier13,Jofre15}, thus encompassing the ${\sim}1.5\,\text{M}_{\sun}$ threshold and calling for a re-evaluation of its fundamental properties.

The rest of the paper is organised as follows. In Sect.~\ref{sec:analysis} we conduct both a spectroscopic and asteroseismic analyses of HD~212771. This is followed in Sect.~\ref{sec:starprop} by the estimation of fundamental stellar properties through a grid-based modelling approach that uses global asteroseismic parameters, complementary spectroscopic data and a parallax-based luminosity as input. In Sect.~\ref{sec:planetevol} we determine the post-main-sequence planetary system evolution based on the newly derived stellar properties. Finally, we discuss our results in Sect.~\ref{sec:discuss}.

\section{Observations and data analysis}\label{sec:analysis}

\subsection{High-resolution spectroscopy}\label{sec:spectroscopy}
We base our spectroscopic analysis on a high-quality FEROS \citep[Fiber-fed Extended Range Optical Spectrograph;][]{FEROS} spectrum retrieved from the ESO archives and obtained on 2011 November 15. It covers the spectral domain $3565$--$9215\:\AA$, with a nominal resolving power $R\!\sim\!48{,}000$. 

The initial data reduction steps (i.e., bias subtraction, flat-field correction, removal of scattered light, order extraction, merging of the orders and wavelength calibration) were carried out with the dedicated instrument pipeline. The spectrum was subsequently put in the laboratory rest frame and the continuum was normalised by fitting low-order cubic splines or Legendre polynomials to the line-free regions. Finally, the telluric features affecting the $^{12}$CN lines around $\sim\!8003\:\AA$ were removed. Our other internal mixing diagnostics (notably $[$\ion{O}{i}$]$ $\lambda6300$ and $^{13}$CN $\lambda8004.7$) are not significantly affected by telluric contamination. These last reduction steps were carried out using standard tasks implemented in the {\sc iraf}\footnote{{\sc iraf} is distributed by the National Optical Astronomy Observatories, which is operated by the Association of Universities for Research in Astronomy, Inc., under cooperative agreement with the National Science Foundation.} software.

The spectral analysis is similar to that carried out for \textit{CoRoT} red giants by \citet{morel14}. Although a number of improvements on the analysis procedure have since been implemented (listed below), this nevertheless leads to small differences (i.e., within the quoted uncertainties) in the estimated atmospheric parameters and elemental abundances: 
\begin{itemize}[leftmargin=5pt,noitemsep,topsep=2pt]
\item Two \ion{Fe}{ii} lines ($\lambda5991$ and $\lambda6416$) were discarded from the analysis because they tend to systematically yield discrepant abundances.
\item A change to the atomic data for some CN features around $\sim\!6707.6\:\AA$ was applied, which leads to a better fit of the blend formed by \ion{Li}{i} $\lambda6708$ and a nearby \ion{Fe}{i} line at $\sim\!6707.4\:\AA$.
\item A better removal of the telluric features affecting the $^{12}$CN lines around $\sim\!8003\:\AA$ (and occasionally $[$\ion{O}{i}$]$ $\lambda6300$) was implemented as well as a more precise assessment of the effect of telluric subtraction on the $^{12}$C/$^{13}$C isotopic ratio.
\end{itemize}

The atmospheric parameters ($T_\text{eff}$, $\log g_\text{spec}$ and microturbulence $\xi$) and abundances of 12 metals (Fe, Na, Mg, Al, Si, Ca, Sc, Ti, Cr, Co, Ni and Ba) are self-consistently determined from the spectrum using a classical curve-of-growth analysis. We computed the relative abundance of $\alpha$ elements as the unweighted mean of the Mg, Si, Ca and Ti abundances, resulting in $[\alpha/\text{Fe}]\!=\!0.06\pm0.02$. The Li, C, N and O abundances (as well as the $^{12}$C/$^{13}$C isotopic ratio) are derived from spectral synthesis of atomic or molecular features. Lithium abundance and the carbon isotopic ratio may be of special interest to some readers, as they carry signatures of extra mixing processes taking place during late stages of stellar evolution \citep[e.g.,][and references therein]{Lagarde12}.

All calculations assume local thermodynamic equilibrium (LTE). Kurucz atmosphere models and the line analysis software {\sc moog}\footnote{\url{http://www.as.utexas.edu/~chris/moog.html}} are used. As discussed by \citet{morel14}, the use of plane-parallel or spherical MARCS model atmospheres does not lead to appreciable differences in GK subgiants at near-solar metallicity. Excitation and ionisation equilibrium of iron are used to derive $T_\text{eff}$ and $\log g_\text{spec}$, while the microturbulence was inferred by requiring no dependence between the \ion{Fe}{i} abundances and the line strength. A total of 55 \ion{Fe}{i} and 8 \ion{Fe}{ii} lines were used. The line-to-line scatter of the Fe abundances is $\sim\!0.035\:\text{dex}$ in both cases.

The results of the spectroscopic analysis are provided in Table \ref{tab:properties1}. $T_\text{eff}$ and $[\text{Fe}/\text{H}]$ will later be used as input in the grid-based modelling (see Sect.~\ref{sec:starprop}). Note that the quoted C, O and Na abundances have not been corrected for their dependence on $[\text{Fe}/\text{H}]$ due to the chemical evolution of the Galaxy \citep[see][]{morel14}. We also provide the non-LTE Li abundance using a correction interpolated from the tables of \citet{lind09}. The abundances based on $[$\ion{O}{i}$]$ $\lambda6300$ are believed to be generally more reliable than those yielded by the \ion{O}{i} triplet because they are not affected by departures from LTE \citep[e.g.,][]{schuler06}. However, in our case both values turn out to be fully consistent.

A number of independent studies have provided atmospheric parameters and elemental abundances for HD~212771 based on high-resolution spectroscopy \citep{discovery,Maldonado13,Mortier13,Santos13,Jofre15}. From these we determined a median iron abundance $[\text{Fe}/\text{H}]\!=\!-0.14$ ($0.04\:\text{dex}$ scatter), a median effective temperature $T_\text{eff}\!=\!5085\:\text{K}$ ($24\:\text{K}$ scatter) and a median surface gravity $\log g_\text{spec}\!=\!3.50$ ($0.09\:\text{dex}$ scatter), in good agreement with the values determined in this work.

\begin{table}
 \centering
 \caption{Atmospheric parameters and elemental abundances.}
 \label{tab:properties1}
 \begin{tabular}{lc}
  \hline
  Parameter & Value \\
  \hline
  $T_\text{eff}$ $[$K$]$ & $5065\pm75$ \\
  $\log g_\text{spec}$ $[$cgs$]$ & $3.37\pm0.17$ \\
  $\xi$ $[$km$\,$s$^{-1}]$ & $1.15\pm0.06$ \\
  \hline
  $[\text{Fe}/\text{H}]$ & $-0.10\pm0.10$ \\
  $[\text{Na}/\text{Fe}]$ & $-0.01\pm0.08$ \\
  $[\text{Mg}/\text{Fe}]$ & $0.04\pm0.10$ \\
  $[\text{Al}/\text{Fe}]$ & $0.08\pm0.10$ \\
  $[\text{Si}/\text{Fe}]$ & $0.10\pm0.08$ \\
  $[\text{Ca}/\text{Fe}]$ & $0.09\pm0.05$ \\
  $[\text{Sc}/\text{Fe}]$ & $0.00\pm0.13$ \\
  $[\text{Ti}/\text{Fe}]$ & $0.02\pm0.09$ \\
  $[\text{Cr}/\text{Fe}]$ & $0.04\pm0.05$ \\
  $[\text{Co}/\text{Fe}]$ & $0.04\pm0.08$ \\
  $[\text{Ni}/\text{Fe}]$ & $-0.02\pm0.05$ \\
  $[\text{Ba}/\text{Fe}]$ & $0.14\pm0.14$ \\
  \hline
  $[\text{Li}/\text{H}]$ LTE & $-0.49\pm0.13$ \\
  $[\text{Li}/\text{H}]$ non-LTE & $-0.40\pm0.13$ \\
  $[\text{C}/\text{Fe}]$ & $-0.11\pm0.09$ \\
  $[\text{N}/\text{Fe}]$ & $0.18\pm0.13$ \\
  $[\text{O}/\text{Fe}]$ & $0.04\pm0.14$ \\
  $^{12}$C/$^{13}$C & $11\pm5$ \\
  \hline
 \end{tabular}
\end{table}

\subsection{Asteroseismology with K2}\label{sec:asteroseismology}
Solar-like oscillations are predominantly standing acoustic waves (or p modes). The oscillation modes are characterised by the radial order $n$, the spherical degree $l$ and the azimuthal order $m$. Radial modes have $l\!=\!0$, whereas non-radial modes have $l\!>\!0$. Observed oscillation modes are typically high-order modes of low spherical degree, with the associated power spectrum showing a pattern of peaks with near-regular frequency spacings. The most prominent separation is the large frequency separation, $\Delta \nu$, between neighbouring overtones with the same spherical degree. The large frequency separation essentially scales as the square root of the mean stellar density, $\langle\rho\rangle^{1/2}$ \citep{Ulrich86,BG94}. Moreover, oscillation mode power is modulated by an envelope that generally assumes a bell-shaped appearance. The frequency at the peak of the power envelope is referred to as the frequency of maximum oscillation amplitude, $\nu_\text{max}$, which has been shown to scale as $g\,T_\text{eff}^{-1/2}$ \citep{Brown91,KB95,Belkacem11}. The fact that $\nu_\text{max}$ mainly depends on $g$ makes it an indicator of the evolutionary state of a star.

\begin{figure*}
    \includegraphics[width=1.5\columnwidth]{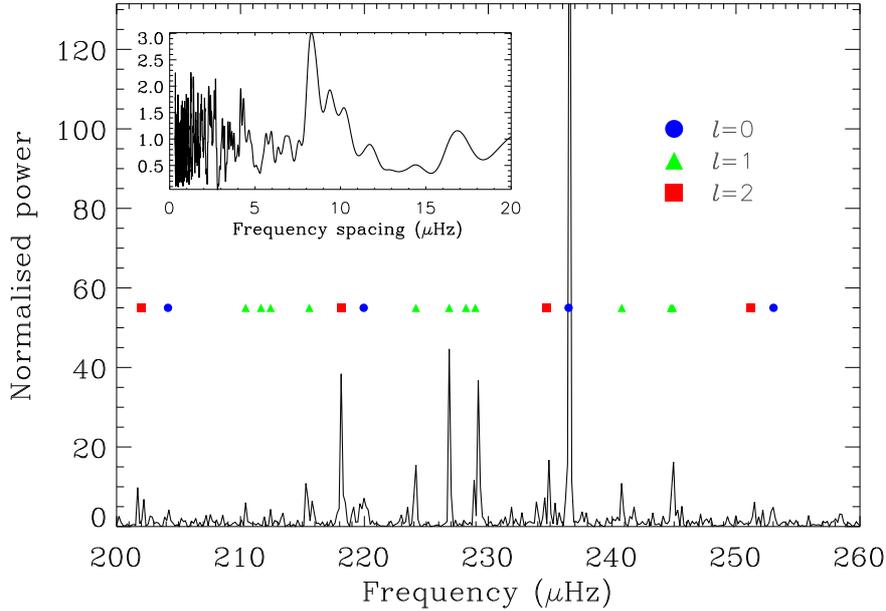}
    \caption{Background-corrected oscillation spectrum of HD~212771. The power spectrum has been normalised after dividing it by a heavily smoothed (using a median filter) version of itself. Radial ($l\!=\!0$) mode frequencies were predicted by applying the universal red-giant oscillation pattern \citep{Mosser11} using the measured $\Delta\nu$ value. Quadrupole ($l\!=\!2$) mode frequencies were visually identified, whereas the locations of dipole ($l\!=\!1$) mixed modes were predicted by adjusting the parameters in the asymptotic expression \citep{Mosser12} until satisfactory agreement was obtained visually. They are represented by blue circles, red squares and green triangles, respectively. The inset shows the power spectrum of the power spectrum (PSPS), computed for the region around $\nu_\text{max}$.}
    \label{fig:powerspec}
\end{figure*}

Substantial changes to the properties of solar-like oscillations occur following exhaustion of hydrogen in the core. The frequencies of non-radial modes, noticeably those of dipole ($l\!=\!1$) modes, are shifted as they undergo avoided crossings, which arise from the coupling between p modes in the outer envelope and g (or gravity) modes trapped in the stellar interior \citep{Osaki75,Aizenman}. At the avoided crossings modes have a mixed nature, with both p- and g-mode behaviour. Ultimately this leads to significant departures from the near-regular spacing in the oscillation spectrum. In contrast with p modes, high-order, low-degree g modes are nearly uniformly spaced in period, not in frequency. The observed period spacing of $l\!=\!1$ mixed modes will be significantly smaller than the underlying asymptotic g-mode period spacing, $\Delta\Pi_1$, as a result of mode bumping. However, provided that a sufficient number of mixed modes are observed, the value of $\Delta\Pi_1$ may be inferred. Stars on the subgiant branch display a strong dependence of $\Delta\Pi_1$ on stellar mass \citep[but also on rotation and convective-core overshooting; e.g.,][]{Benomar12,retiredKpl,Lagarde16}, thus making this quantity particularly useful in the estimation of mass.

HD~212771 was observed by K2 in short-cadence mode (with cadence $\Delta t\!\sim\!58.85\:\text{s}$) during its Campaign 3 (C3) as part of Guest Observer Programme GO3025 (PI: Johnson), spanning a total of approximately 69.2 days. Although C3 had a nominal duration of 80 days, the campaign ended earlier than expected because the on-board storage filled up faster than anticipated due to unusually poor data compression\footnote{\url{https://keplerscience.arc.nasa.gov/k2-data-release-notes.html\#k2-campaign-3}}. We used the {\sc k2p$^2$} pipeline \citep{K2P2} --- taking the reprocessed short-cadence target pixel data from Data Release 10 as input --- to prepare a light curve for asteroseismic analysis, after which additional corrections were made using the filtering described by \citet{HandLund}. A least-squares sine-wave-fitting method was then used to compute the rms-scaled power spectrum of the light curve.

The global asteroseismic parameters $\Delta \nu$ and $\nu_\text{max}$ were measured based on the analysis of the above power spectrum. Figure \ref{fig:powerspec} shows the background-corrected power spectrum over the frequency range occupied by the oscillations. The inset shows the power spectrum of the power spectrum (PSPS), computed for the region around $\nu_\text{max}$. The prominent peak in the PSPS lies at $\Delta\nu/2$, a clear signature of the near-regular spacing in the oscillation spectrum. A complementary range of well-tested automated methods was used in the analysis, which had previously been extensively applied to data from the nominal \textit{Kepler} mission \citep[e.g.,][]{Huber09,Campante10,Hekker10,Campante12,DaviesMiglio}. We adopted the values of $\Delta \nu$ and $\nu_\text{max}$ returned by the method described in \citet{Huber09}, with final uncertainties recalculated by adding in quadrature the corresponding formal uncertainty and the standard deviation of the parameter estimates given by all methods. The consolidated values are then $\Delta \nu\!=\!16.5\pm0.3\:\mu\text{Hz}$ and $\nu_\text{max}\!=\!231\pm3\:\mu\text{Hz}$, i.e., to better than 2 per cent precision in both parameters \citep[cf.][]{Hekker11,DaviesMiglio}. Although a measurement of $\Delta\Pi_1$ has been attempted, the small number of observed mixed modes per radial order hindered estimation of a precise asymptotic g-mode period spacing.

\section{Estimation of fundamental stellar properties}\label{sec:starprop}
Fundamental stellar properties can be estimated by comparing global asteroseismic parameters and complementary spectroscopic data to the outputs of stellar evolutionary models. This is often done following a grid-based approach, whereby observables are matched to well-sampled grids of stellar evolutionary tracks \citep[e.g.,][]{Stello09,Basu10,Basu12,ChaplinSci,Creevey12}. Here, we employ the Bayesian code {\sc param} \citep{PARAM1,PARAM2,Rodrigues17}. Based on a given set of observables, the code first computes the probability density functions (PDFs) for the stellar mass, $M$, radius, $R$, age, $t$, surface gravity, $\log g$, and mean density, $\log (\rho/\rho_{\sun})$, as well as for the absolute magnitudes in a number of widely used bandpasses. In a second step, the code combines apparent and absolute magnitudes to derive the distance to the star, $d$. We note that, in the latest version of {\sc param} \citep[][]{Rodrigues17}, $\Delta\nu$ is not assumed to follow a simple scaling relation with $\langle\rho\rangle^{1/2}$, but instead it is based on theoretical radial-mode frequencies computed for the models in the grid.

The underlying grid of stellar evolutionary tracks on which this latest version of {\sc param} runs has been computed using the Modules for Experiments in Stellar Astrophysics \citep[{\sc mesa};][]{Paxton11,Paxton13} evolution code. The relevant input physics is summarised next \citep[see also][]{Rodrigues17}: The \citet{GN93} heavy elements partition was adopted. The {\sc opal} equation of state \citep{RN02} and opacities \citep{IR96} were used, complemented at low temperatures by opacities from \citet{Ferguson05}. Nuclear reaction rates were obtained from tables provided by the NACRE collaboration \citep{Angulo99}. The atmosphere model follows \citet{KS66}. The mixing length theory was used to describe convection (a solar-calibrated parameter $\alpha_\text{MLT}\!=\!1.9657$ was adopted). Overshooting was included during the core-convective burning phases according to the \citet{Maeder75} step function scheme. The extent of convective-core overshooting during the main sequence was taken as $\alpha_\text{ov}\!=\!0.2\,H_p$, where $H_p$ is the pressure scale height (i.e., the radial distance over which the pressure changes by a factor of e) at the boundary of the convective core. No diffusion, mass loss or effects of rotational mixing have been included.

\begin{figure}
    \includegraphics[width=\columnwidth]{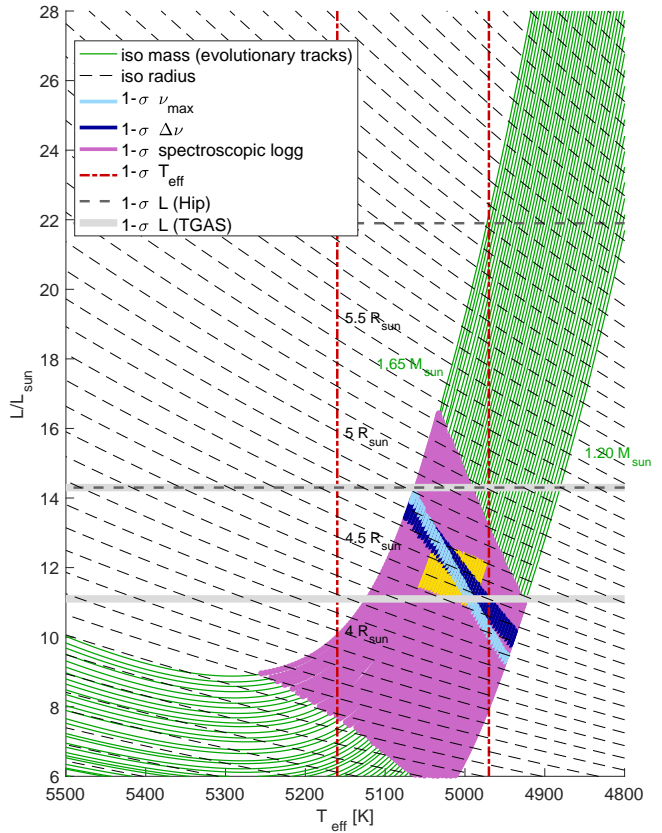}
    \caption{Hertzsprung--Russell diagram showing observational constraints used in the estimation of fundamental stellar properties for HD~212771. Stellar evolutionary tracks spanning the mass range $1.20$--$1.65\,\text{M}_{\sun}$ (in steps of $0.01\,\text{M}_{\sun}$) were computed using {\sc mesa} for a fixed metallicity ($[\text{Fe}/\text{H}]\!=\!-0.10$) and are shown as solid green lines. Contours of constant stellar radius are represented by dashed lines. Coloured bands represent the $1$-$\sigma$ observational constraints on $\Delta\nu$, $\nu_\text{max}$ and $\log g_\text{spec}$, whilst $1$-$\sigma$ lower and upper bounds on $T_\text{eff}$, and on both the \textit{Gaia} DR1 (`TGAS') and \textit{Hipparcos} (`Hip') parallax-based luminosities are indicated by different line styles/colours. The yellow-shaded box represents the $68\,\%$ Bayesian credible region for stellar mass and radius corresponding to the reference solution (see Table \ref{tab:properties2}).}
    \label{fig:HR}
\end{figure}

The primary set of observables consists of $\left\{[\text{m}/\text{H}],T_\text{eff},\Delta\nu,\nu_\text{max},L\right\}$, where $[\text{m}/\text{H}]$ is the overall metallicity and $L$ is the stellar luminosity. A parallax-based luminosity (see below) is used here as an additional independent constraint. Given the modest $\alpha$ enhancement (see Sect.~\ref{sec:spectroscopy}), we therefore assume $[\text{m}/\text{H}]\!\approx\![\text{Fe}/\text{H}]$. Furthermore, contributions of $0.062\:\text{dex}$ in $[\text{m}/\text{H}]$ and $59\:\text{K}$ in $T_\text{eff}$ were added in quadrature to the formal uncertainties to account for systematic differences between spectroscopic methods \citep{Torres12}, producing $[\text{m}/\text{H}]\!=\!-0.10\pm0.12$ and $T_\text{eff}\!=\!5065\pm95\:\text{K}$. The stellar luminosity was calculated via the parallax, $\pi\!=\!8.95\pm0.23\:\text{mas}$, given in\footnote{\textit{Gaia} DR1 incorporates earlier positional information through the Tycho-\textit{Gaia} astrometric solution \citep[TGAS;][]{Michalik15}. We note that a systematic component of $0.3\:\text{mas}$ was added to the formal parallax uncertainty \citep[][]{BrownGaia} prior to its use in Eq.~(\ref{eq:lum}).} \textit{Gaia} Data Release 1 \citep[\textit{Gaia} DR1;][]{GaiaDR1} using the following relation from \citet{Pijpers03}:
\begin{multline}
\label{eq:lum}
\log(L/\text{L}_{\sun}) = 4.0 + 0.4M_{\text{bol},\sun} - 2.0\log\pi[\text{mas}] \\
- 0.4(V - A_V + \text{BC}_V) \, ,
\end{multline}  
where we have adopted $M_{\text{bol},\sun}\!=\!4.73\:\text{mag}$ \citep{Torres10} for the solar bolometric magnitude, $A_V\!=\!0.11\pm0.02\:\text{mag}$ is the extinction in the $V$ band and $\text{BC}_V\!=\!-0.28\:\text{mag}$ is the bolometric correction from the \citet{Flower96} polynomials presented in \citet{Torres10}, which use $T_\text{eff}$ as input. The extinction was estimated using the stellar coordinates and parallax as input to the code {\sc mwdust}\footnote{\url{https://github.com/jobovy/mwdust}} \citep{Bovy15}, for which we adopted the \citet{Green15} dust map. A value of the luminosity based on the \textit{Hipparcos} parallax \citep[$\pi\!=\!7.63\pm0.80\:\text{mas}$;][]{Hipparcos} was also calculated and {\sc param} run for a second time (now using $A_V\!=\!0.15\pm0.01\:\text{mag}$). Finally, in order to derive the distance to the star, we also made use of the SDSS $griz$ and 2MASS $JHK_\text{s}$ apparent magnitudes as provided in the K2 Ecliptic Plane Input Catalog \citep[EPIC;][]{EPIC}.

\begin{figure*}
    \includegraphics[width=1.5\columnwidth]{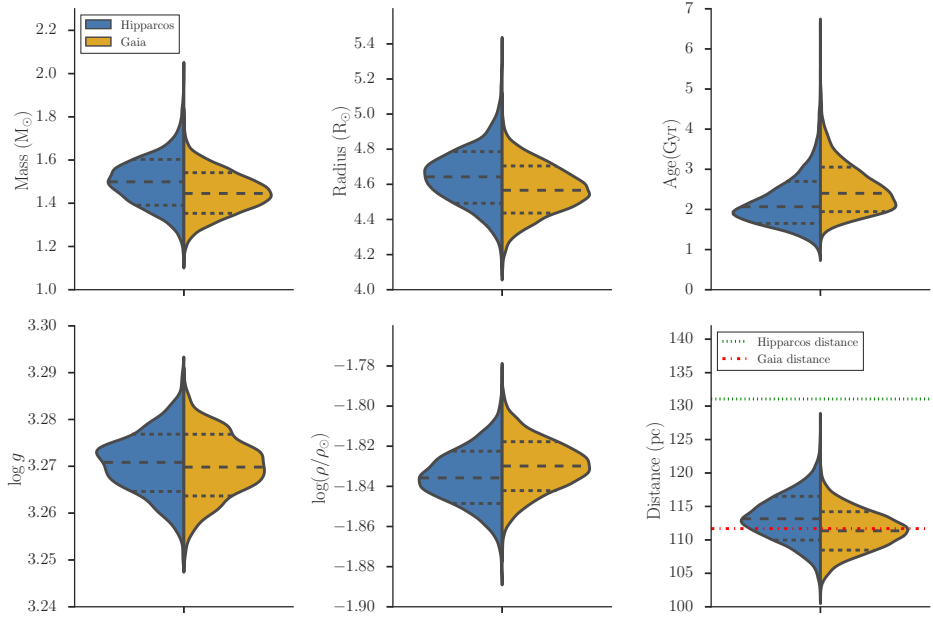}
    \caption{Probability density functions of the fundamental stellar properties obtained using {\sc param}. These are displayed as a series of violin plots, with a rotated kernel density estimate of the PDF (or PDFs in this case) on each side. Here, we plot the PDFs obtained when considering either the \textit{Gaia} DR1 or \textit{Hipparcos} parallax-based luminosities as input. The corresponding median and $68\,\%$ Bayesian credible region are shown under the PDFs. The \textit{Gaia} DR1 and \textit{Hipparcos} parallax-based distances (i.e., $1/\pi[\text{arcsec}]$) are indicated in the bottom rightmost panel as horizontal lines.}
    \label{fig:violin}
\end{figure*}

The H--R diagram in Fig.~\ref{fig:HR} shows the several observational constraints used in the analysis (as well as that corresponding to $\log g_\text{spec}$) plotted atop a sequence of stellar evolutionary tracks of fixed metallicity. This figure serves to illustrate the constraining power gained --- and consequent reduction in model degeneracy --- when using $\Delta\nu$ and $\nu_\text{max}$ in addition to the use of classical spectroscopic constraints and a parallax-based luminosity. We note that while the \textit{Gaia} DR1 parallax-based luminosity is consistent (at the $1$-$\sigma$ level) with the constraints imposed by asteroseismology, the same is not true with the \textit{Hipparcos} parallax-based luminosity.

Figure \ref{fig:violin} shows the output of our analysis as a series of violin plots (see also Table \ref{tab:properties2}). These plots show the outcome of using either the \textit{Gaia} DR1 or \textit{Hipparcos} parallax-based luminosities as input to {\sc param}. Despite the discrepant (at the $1$-$\sigma$ level) luminosity constraints (see Fig.~\ref{fig:HR}), both solutions lead to consistent PDFs for the stellar properties, an indication that the output produced by {\sc param} is being dominated by the asteroseismic constraints. Moreover, the derived distances, obtained by combining apparent and absolute magnitudes, are consistent with the \textit{Gaia} DR1 parallax-based distance (see bottom rightmost panel of Fig.~\ref{fig:violin}). Therefore, we henceforth adopt as the reference solution the set of fundamental stellar properties obtained using the \textit{Gaia} DR1 parallax-based luminosity as input to {\sc param}.

The derived $\log g$ is lower than $\log g_\text{spec}$ (though still consistent at the $1$-$\sigma$ level given the large uncertainty affecting the latter value) and the median literature value (see Sect.~\ref{sec:spectroscopy}). Spectroscopic $\log g$ values tend to be systematically higher than asteroseismic values for red giants, as demonstrated for field red giants with APOGEE $H$-band spectra \citep{Pinsonneault14} and for evolved \textit{Kepler} exoplanet-host stars with optical high-resolution spectra \citep{Huber13}. Such an offset, which is largely dependent on the stellar evolutionary state, as well as on stellar mass in the case of low-luminosity red giants, is believed to be inherent to the spectroscopic analysis \citep{Masseron}.

Furthermore, the reference asteroseismic mass\footnote{The (model-independent) mass obtained from asteroseismic scaling relations ($1.54\pm0.06\:\text{M}_{\sun}$) is consistent with the reference asteroseismic mass at the $1$-$\sigma$ level. A correction to the $\Delta \nu$ scaling relation, which can amount to as much as 2 per cent (i.e., $f_{\Delta \nu}\la1.02$) for the current mass, evolutionary state and chemical composition \citep[e.g.,][]{Sharma16,Rodrigues17}, translating into a correction $f_{\Delta \nu}^{-4}\ga0.92$ to the stellar mass from scaling relations, would only bring it closer to the reference asteroseismic value.} ($1.45^{+0.10}_{-0.09}\:\text{M}_{\sun}$; see Table \ref{tab:properties2}) places HD~212771 in the retired A star category (its main-sequence progenitor was likely an early F-type star). This mass is consistent with spectroscopy-based masses found in the literature, namely, those derived by\footnote{Note that \citet{Mortier13} give preference to stellar masses derived using the TS13--SO08 line list (their table 5).} \citet{Mortier13} ($1.51\pm0.08\:\text{M}_{\sun}$) and \citet{Jofre15} ($1.60\pm0.13\:\text{M}_{\sun}$), while being significantly higher than the value reported in the discovery paper \citep[$1.15\pm0.08\:\text{M}_{\sun}$;][]{discovery}.

We tested the robustness of the reference asteroseismic mass by evaluating the impact that different input physics has on its estimation. Varying the convective-core overshooting parameter (i.e., $\alpha_\text{ov}\!=\!\{0,0.1\,H_p,0.2\,H_p\}$) in the underlying {\sc mesa} grid led to no significant variation of the estimated asteroseismic mass. This parameter is especially relevant, as stars in the mass range under study develop convective cores while on the main sequence (i.e., for $M\ga1.1\,\text{M}_{\sun}$ at solar metallicity). Alternatively, we used an earlier version of {\sc param} \citep{PARAM2} that runs on a grid of stellar evolutionary tracks computed using the PAdova \& TRieste Stellar Evolution Code \citep[{\sc parsec};][]{Bressan12}. In particular, the prescription for convective-core overshooting on the main sequence implemented in {\sc parsec} differs from that in {\sc mesa} in that $\alpha_\text{ov}$ increases with stellar mass, as opposed to being set to a fixed fraction of $H_p$ \citep[a detailed comparison between the input physics in both grids can be found in][]{Bossini15}. Again, we saw no significant variation of the estimated asteroseismic mass.

The fact that the reference asteroseismic mass significantly differs from the value reported in the discovery paper, which has been used to inform the stellar mass-planet occurrence relation, is nonetheless a reason for concern. To delve into the possible cause(s) for this discrepancy, we decided to remove the asteroseismic constraints altogether --- thus using only classical spectroscopic constraints and a parallax-based luminosity in our analysis (i.e., $\left\{[\text{m}/\text{H}],T_\text{eff},L\right\}$) --- and assess how susceptible a spectroscopy-based mass is to different input physics. Use of the reference {\sc mesa} grid (i.e., with $\alpha_\text{ov}\!=\!0.2\,H_p$) leads to a mass estimate ($1.41^{+0.17}_{-0.21}\:\text{M}_{\sun}$) consistent with the reference asteroseismic value. Notice, however, the appreciable loss in relative precision of the estimated mass as a result of the lack of asteroseismic information \citep[cf.][]{Rodrigues17}. Moreover, no significant variation of the estimated mass is seen when varying $\alpha_\text{ov}$ in the underlying {\sc mesa} grid. Use of the {\sc parsec} grid\footnote{Incidentally, the treatment of convective-core overshooting implemented in {\sc parsec} is similar to that adopted in the {\sc y$^2$} tracks \citep{Y2} used to derive stellar properties in the discovery paper (i.e., $\alpha_\text{ov}$ increases with stellar mass).}, on the other hand, leads to a mass estimate ($1.26^{+0.13}_{-0.16}\:\text{M}_{\sun}$) that, although marginally compatible with the reference asteroseismic value, is now fully consistent with the value reported in the discovery paper. The change in the mass estimate comes from a subtle combination of the various differences in input physics between {\sc mesa} and {\sc parsec}, and cannot be attributed solely to a different treatment of convective-core overshooting. We further tested the robustness of the spectroscopy-based mass by evaluating the impact of potential biases in the spectroscopic parameters while adopting fixed input physics (the reference {\sc mesa} grid was used). This was done by artificially introducing small biases (of magnitude $1\sigma$) in the observables $[\text{m}/\text{H}]$ and $T_\text{eff}$, having perturbed one property at a time. The impact of using deflated uncertainties on these observables was then also tested (e.g., for the metallicity, an uncertainty of $0.03\:\text{dex}$, as provided in the discovery paper, was used). As with the input physics, the introduction of a bias in metallicity (coupled to an underestimation of its uncertainty) can lead to noticeable excursions of the estimated mass.

\begin{table}
 \centering
 \caption{Fundamental stellar properties (using either the \textit{Gaia} DR1 or \textit{Hipparcos} parallax-based luminosities as input to {\sc param}).}
 \label{tab:properties2}
 \begin{tabular}{lcc}
  \hline
  Parameter$^a$ & Value$^b$ (\textit{Gaia}) & Value (\textit{Hipparcos})\\
  \hline
  $M$ $[\text{M}_{\sun}]$ & $1.45^{+0.10\,(0.18)}_{-0.09\,(0.18)}$ & $1.50^{+0.10\,(0.23)}_{-0.11\,(0.21)}$ \\[0.08in]
  $L$ $[\text{L}_{\sun}]$ & $12.7\pm1.6$ & $18.1\pm3.8$ \\[0.08in]
  $R$ $[R_{\sun}]$ & $4.55^{+0.14\,(0.26)}_{-0.13\,(0.26)}$ & $4.62^{+0.14\,(0.31)}_{-0.15\,(0.30)}$ \\[0.08in]
  $t$ $[$Gyr$]$ & $2.52^{+0.69\,(1.56)}_{-0.52\,(0.86)}$ & $2.18^{+0.70\,(1.63)}_{-0.45\,(0.80)}$ \\[0.08in]
  $\log g$ $[$cgs$]$ & $3.267^{+0.006\,(0.012)}_{-0.006\,(0.013)}$ & $3.268^{+0.006\,(0.012)}_{-0.006\,(0.013)}$ \\[0.08in]
  $\log (\rho/\rho_{\sun})$ & $-1.831^{+0.012\,(0.024)}_{-0.012\,(0.024)}$ & $-1.837^{+0.013\,(0.026)}_{-0.013\,(0.027)}$ \\[0.08in]
  $d$ $[$pc$]$ & $111.3^{+2.8\,(5.7)}_{-2.8\,(5.6)}$ & $113.2^{+3.3\,(6.4)}_{-3.2\,(6.3)}$ \\
  \hline
  \multicolumn{3}{l}{$^a$ The $68\,\%$ ($95\,\%$) Bayesian credible region is given outside}\\ 
  \multicolumn{3}{l}{(inside) brackets. For $L$, the $1$-$\sigma$ error is given.}\\
  \multicolumn{3}{l}{$^b$ Adopted as the reference solution.}\\
 \end{tabular}
\end{table}

\section{Post-main-sequence planetary system evolution}\label{sec:planetevol}

Implications for the fates of planetary systems are sensitively dependent on stellar mass because that determines (i) the spatial extent within which all planetary material will be engulfed through tides, (ii) the possibility of gravitational scattering amongst the surviving bodies due to the amount of stellar mass lost, and (iii) the capability of the star's luminosity to break up sub-planet-sized objects in asteroid-belt and Kuiper-belt analogues through rotational fission.  All these components dictate the planetary system architecture around the eventual white dwarf \citep{veras2016a}.

The current population of white dwarfs in fact primarily arises from A-star progenitors \citep{treetal2016}, which highlights the importance of accurately determining their masses during the subgiant/giant branch for determining the fate of planetary systems \citep{veretal2016b}. Based on the reference asteroseismic mass derived in the previous section, we go on in this section to predict the evolution of the Jovian planet orbiting HD~212771 under the effects of tidal forces and stellar mass loss.

We performed a simulation of the future evolution of HD~212771 by assuming $M\!=\!1.45\,\text{M}_{\sun}$ and metallicity $Z\!=\!0.01$ through the use of the {\sc sse} code \citep{huretal2000}. HD~212771 will achieve its maximum radius of $1.32\:\text{au}$ at $1.28\:\text{Myr}$ into the asymptotic-giant-branch phase. At this time, the star will have lost $0.61\,\text{M}_{\sun}$ of its mass. The star will then continue to lose another $0.25\,\text{M}_{\sun}$ to achieve a final white dwarf mass of $0.59\:\text{M}_{\sun}$. Although mass loss pushes the planet outward, this process competes with tides induced on the planet by the star. The outcome is dependent on both the planet's orbital and physical properties. We estimate that the planet has a semimajor axis $a\!\sim\!1.15\:\text{au}$ from our knowledge of its orbital period and minimum mass, as well as the stellar mass from asteroseismology. Given an estimate of HD~212771~b's eccentricity \citep[$e\!=\!0.11$;][]{discovery}, its pericentre is then about $1.02\:\text{au}$. 

These orbital and physical properties ultimately imply that the planet will not survive the star's post-main-sequence evolution and be engulfed as the star ascends the asymptotic giant branch. The planet's semimajor axis dictates that its orbit will expand adiabatically, with a semimajor axis increase proportional to stellar mass loss decrease and no change in eccentricity \citep{veretal2011}. Hence, at the time of maximum radial extent, the planet will have been pushed out to a supposedly safe distance of $1.63\:\text{au}$. However, tidal effects extend beyond the reach of the stellar surface, a fact not appreciated by some earlier studies \citep{sacetal1993}. This extra reach is not trivially computed along the asymptotic giant branch and is dependent on many assumptions, including the frequency and magnitude of thermal pulses, and stellar spin \citep{musvil2012,adablo2013,norspi2013}. The top panel of fig.~2 of \citet{musvil2012} demonstrates that HD~212771~b has no chance of surviving: Jupiter-like planets are much more strongly affected by tidal interactions with asymptotic-giant-branch stars than Earth-like planets (bottom panel of that same figure). However, even if HD~212771~b was an Earth-like planet, that figure illustrates that the planet would be unlikely to survive. Once tides dominate over mass loss, the actual inspiral timescale will be very short, on the order of decades \citep{staetal2016}.

\section{Discussion}\label{sec:discuss}
\citet{Lloyd11} hypothesised that the masses of subgiant and low-luminosity red-giant stars targeted by Doppler-based planet surveys --- typically derived from a combination of spectroscopy  and isochrone fitting --- may be systematically overestimated. In this study we tested this hypothesis for the particular case of the exoplanet-host star HD~212771, for which asteroseismology with K2 has recently been made possible. 

Stringent tests of the accuracy of asteroseismic masses of stars in similar evolutionary states to HD~212771 have shown no evidence of systematic offsets \citep[e.g.,][and references therein]{Miglio16,Sharma16,Stello16}. Crucially, such tests employ stellar-model-based corrections to the $\Delta\nu$ scaling relation. The use of asteroseismic scaling relations at face value is, on the other hand, expected to lead to systematic offsets of about 10 per cent in the determination of mass, as shown, e.g., by \citet{Miglio16} and \citet{Sharma16}, and with higher significance by \citet{Gaulme16}. We estimated the mass of HD~212771 through a grid-based modelling approach that uses global asteroseismic parameters, complementary spectroscopic data and a parallax-based luminosity as input. Importantly, the mass of HD~212771 was estimated by comparing the observed $\Delta\nu$ to a value computed from theoretical radial-mode frequencies and not from asteroseismic scaling relations (see Sect.~\ref{sec:starprop}).

The reference asteroseismic mass ($1.45^{+0.10}_{-0.09}\:\text{M}_{\sun}$; see Table \ref{tab:properties2}) is consistent with spectroscopy-based masses found in the literature, namely, those derived by \citet{Mortier13} ($1.51\pm0.08\:\text{M}_{\sun}$) and \citet{Jofre15} ($1.60\pm0.13\:\text{M}_{\sun}$). This estimate is nonetheless significantly higher than the value reported in the discovery paper \citep[$1.15\pm0.08\:\text{M}_{\sun}$;][]{discovery}, which has been used to inform the stellar mass-planet occurrence relation. Having established the robustness of the reference asteroseismic mass in Sect.~\ref{sec:starprop}, we attribute this discrepancy to the susceptibility of spectroscopy-based estimates to different input physics and to potential biases in metallicity (coupled to an underestimation of its uncertainty).

The newly derived stellar mass from asteroseismology hence places HD~212771 in the retired A star category. Furthermore, this result does not lend support to the hypothesis put forward by \citet{Lloyd11}. If this were to be systematic, i.e., if the masses of evolved stars targeted by Doppler-based planet surveys were instead being underestimated, then the stellar mass-planet occurrence relation would turn out to be less steep than currently thought. An alternative interpretation, however, has to do with the notion that the uncertainties on the masses of retired A stars found in the literature are being significantly underestimated (as a result of the underestimation of the uncertainties on spectroscopic parameters). \citet{Schlaufman13} showed that the large Galactic space motions of subgiant host stars require that on average their masses be similar to those of main-sequence F- and G-type hosts. Therefore, if the masses of retired A stars were to be characterised by a scatter a few times the nominal mass uncertainty (even in the absence of a systematic bias), there could be enough contamination in the sample from low-mass stars to explain the larger-than-expected space motions of subgiant host stars. Presently, asteroseismic masses are available for only two subgiant/low-luminosity red-giant stars targeted by Doppler-based planet surveys. Besides HD~212771 (this work), for which there is a $3.8$-$\sigma$ discrepancy between the discovery mass and the asteroseismic mass, the mass of the (non-host) retired A star HD~185351 \citep[$1.87\pm0.07\:\text{M}_{\sun}$;][]{Johnson14} is $4.1$-$\sigma$ higher than its asteroseismic mass \citep[$1.58^{+0.04}_{-0.02}\:\text{M}_{\sun}$;][]{retiredKpl}. Note that the asteroseismic mass is the lower of the two values in the latter case. This might thus be hinting at a level of scatter that is a few times the nominal mass uncertainty.

To be in a position to draw more definitive conclusions, a study is underway of an ensemble of retired A stars with \textit{Kepler}/K2 asteroseismology that have previously been targeted by Doppler-based planet surveys \citetext{T.~S.~H.~North et al., in prep.}. This ensemble will be greatly expanded once asteroseismology with the \textit{Transiting Exoplanet Survey Satellite} \citep[\textit{TESS};][]{TESS} becomes available, as subgiant and low-luminosity red-giant stars will make up a significant fraction of the core asteroseismic targets of that mission \citep{CampanteTESS}.

\section*{Acknowledgements}
This paper includes data collected by the K2 mission. Funding for the K2 mission is provided by the NASA Science Mission directorate. The authors acknowledge the support of the UK Science and Technology Facilities Council (STFC). Funding for the Stellar Astrophysics Centre is provided by The Danish National Research Foundation (Grant DNRF106). DV received funding from the European Research Council under the European Union's Seventh Framework Programme (FP/2007--2013)/ERC Grant Agreement n.~320964 (WDTracer). TM acknowledges financial support from Belspo for contract PRODEX GAIA-DPAC. DH acknowledges support by the Australian Research Council's Discovery Projects funding scheme (project number DE140101364) and support by the National Aeronautics and Space Administration under Grant NNX14AB92G issued through the \textit{Kepler} Participating Scientist Program. MNL acknowledges the support of The Danish Council for Independent Research | Natural Science (Grant DFF-4181-0415). TSR acknowledges the support from CNPq-Brazil and PRIN INAF 2014 - CRA 1.05.01.94.05. This research has made use of the SIMBAD database, operated at CDS, Strasbourg, France.



\bibliographystyle{mnras}
\bibliography{biblio}

\begin{thebibliography}{}
\makeatletter
\relax
\def\mn@urlcharsother{\let\do\@makeother \do\$\do\&\do\#\do\^\do\_\do\%\do\~}
\def\mn@doi{\begingroup\mn@urlcharsother \@ifnextchar [ {\mn@doi@}
  {\mn@doi@[]}}
\def\mn@doi@[#1]#2{\def\@tempa{#1}\ifx\@tempa\@empty \href
  {http://dx.doi.org/#2} {doi:#2}\else \href {http://dx.doi.org/#2} {#1}\fi
  \endgroup}
\def\mn@eprint#1#2{\mn@eprint@#1:#2::\@nil}
\def\mn@eprint@arXiv#1{\href {http://arxiv.org/abs/#1} {{\tt arXiv:#1}}}
\def\mn@eprint@dblp#1{\href {http://dblp.uni-trier.de/rec/bibtex/#1.xml}
  {dblp:#1}}
\def\mn@eprint@#1:#2:#3:#4\@nil{\def\@tempa {#1}\def\@tempb {#2}\def\@tempc
  {#3}\ifx \@tempc \@empty \let \@tempc \@tempb \let \@tempb \@tempa \fi \ifx
  \@tempb \@empty \def\@tempb {arXiv}\fi \@ifundefined
  {mn@eprint@\@tempb}{\@tempb:\@tempc}{\expandafter \expandafter \csname
  mn@eprint@\@tempb\endcsname \expandafter{\@tempc}}}

\bibitem[\protect\citeauthoryear{{Adams} \& {Bloch}}{{Adams} \&
  {Bloch}}{2013}]{adablo2013}
{Adams} F.~C.,  {Bloch} A.~M.,  2013, \mn@doi [\apjl]
  {10.1088/2041-8205/777/2/L30}, \href
  {http://ukads.nottingham.ac.uk/abs/2013ApJ...777L..30A} {777, L30}

\bibitem[\protect\citeauthoryear{{Aizenman}, {Smeyers}  \&
  {Weigert}}{{Aizenman} et~al.}{1977}]{Aizenman}
{Aizenman} M.,  {Smeyers} P.,   {Weigert} A.,  1977, \aap, \href
  {http://adsabs.harvard.edu/abs/1977A%26A....58...41A} {58, 41}

\bibitem[\protect\citeauthoryear{{Angulo} et~al.,}{{Angulo}
  et~al.}{1999}]{Angulo99}
{Angulo} C.,  et~al., 1999, \mn@doi [Nuclear Physics A]
  {10.1016/S0375-9474(99)00030-5}, \href
  {http://adsabs.harvard.edu/abs/1999NuPhA.656....3A} {656, 3}

\bibitem[\protect\citeauthoryear{{Basu}, {Chaplin}  \& {Elsworth}}{{Basu}
  et~al.}{2010}]{Basu10}
{Basu} S.,  {Chaplin} W.~J.,   {Elsworth} Y.,  2010, \mn@doi [\apj]
  {10.1088/0004-637X/710/2/1596}, \href
  {http://adsabs.harvard.edu/abs/2010ApJ...710.1596B} {710, 1596}

\bibitem[\protect\citeauthoryear{{Basu}, {Verner}, {Chaplin}  \&
  {Elsworth}}{{Basu} et~al.}{2012}]{Basu12}
{Basu} S.,  {Verner} G.~A.,  {Chaplin} W.~J.,   {Elsworth} Y.,  2012, \mn@doi
  [\apj] {10.1088/0004-637X/746/1/76}, \href
  {http://adsabs.harvard.edu/abs/2012ApJ...746...76B} {746, 76}

\bibitem[\protect\citeauthoryear{{Belkacem}, {Goupil}, {Dupret}, {Samadi},
  {Baudin}, {Noels}  \& {Mosser}}{{Belkacem} et~al.}{2011}]{Belkacem11}
{Belkacem} K.,  {Goupil} M.~J.,  {Dupret} M.~A.,  {Samadi} R.,  {Baudin} F.,
  {Noels} A.,   {Mosser} B.,  2011, \mn@doi [\aap]
  {10.1051/0004-6361/201116490}, \href
  {http://adsabs.harvard.edu/abs/2011A%26A...530A.142B} {530, A142}

\bibitem[\protect\citeauthoryear{{Benomar}, {Bedding}, {Stello}, {Deheuvels},
  {White}  \& {Christensen-Dalsgaard}}{{Benomar} et~al.}{2012}]{Benomar12}
{Benomar} O.,  {Bedding} T.~R.,  {Stello} D.,  {Deheuvels} S.,  {White} T.~R.,
   {Christensen-Dalsgaard} J.,  2012, \mn@doi [\apjl]
  {10.1088/2041-8205/745/2/L33}, \href
  {http://adsabs.harvard.edu/abs/2012ApJ...745L..33B} {745, L33}

\bibitem[\protect\citeauthoryear{{Bossini} et~al.,}{{Bossini}
  et~al.}{2015}]{Bossini15}
{Bossini} D.,  et~al., 2015, \mn@doi [\mnras] {10.1093/mnras/stv1738}, \href
  {http://adsabs.harvard.edu/abs/2015MNRAS.453.2290B} {453, 2290}

\bibitem[\protect\citeauthoryear{{Bovy}, {Rix}, {Green}, {Schlafly}  \&
  {Finkbeiner}}{{Bovy} et~al.}{2016}]{Bovy15}
{Bovy} J.,  {Rix} H.-W.,  {Green} G.~M.,  {Schlafly} E.~F.,   {Finkbeiner}
  D.~P.,  2016, \mn@doi [\apj] {10.3847/0004-637X/818/2/130}, \href
  {http://adsabs.harvard.edu/abs/2016ApJ...818..130B} {818, 130}

\bibitem[\protect\citeauthoryear{{Bowler} et~al.,}{{Bowler}
  et~al.}{2010}]{Bowler10}
{Bowler} B.~P.,  et~al., 2010, \mn@doi [\apj] {10.1088/0004-637X/709/1/396},
  \href {http://adsabs.harvard.edu/abs/2010ApJ...709..396B} {709, 396}

\bibitem[\protect\citeauthoryear{{Bressan}, {Marigo}, {Girardi}, {Salasnich},
  {Dal Cero}, {Rubele}  \& {Nanni}}{{Bressan} et~al.}{2012}]{Bressan12}
{Bressan} A.,  {Marigo} P.,  {Girardi} L.,  {Salasnich} B.,  {Dal Cero} C.,
  {Rubele} S.,   {Nanni} A.,  2012, \mn@doi [\mnras]
  {10.1111/j.1365-2966.2012.21948.x}, \href
  {http://adsabs.harvard.edu/abs/2012MNRAS.427..127B} {427, 127}

\bibitem[\protect\citeauthoryear{{Brown} \& {Gilliland}}{{Brown} \&
  {Gilliland}}{1994}]{BG94}
{Brown} T.~M.,  {Gilliland} R.~L.,  1994, \mn@doi [\araa]
  {10.1146/annurev.aa.32.090194.000345}, \href
  {http://adsabs.harvard.edu/abs/1994ARA%26A..32...37B} {32, 37}

\bibitem[\protect\citeauthoryear{{Brown}, {Gilliland}, {Noyes}  \&
  {Ramsey}}{{Brown} et~al.}{1991}]{Brown91}
{Brown} T.~M.,  {Gilliland} R.~L.,  {Noyes} R.~W.,   {Ramsey} L.~W.,  1991,
  \mn@doi [\apj] {10.1086/169725}, \href
  {http://adsabs.harvard.edu/abs/1991ApJ...368..599B} {368, 599}

\bibitem[\protect\citeauthoryear{{Campante}}{{Campante}}{2012}]{Campante12}
{Campante} T.~L.,  2012, PhD thesis, Universidade do Porto

\bibitem[\protect\citeauthoryear{{Campante}, {Karoff}, {Chaplin}, {Elsworth},
  {Handberg}  \& {Hekker}}{{Campante} et~al.}{2010}]{Campante10}
{Campante} T.~L.,  {Karoff} C.,  {Chaplin} W.~J.,  {Elsworth} Y.~P.,
  {Handberg} R.,   {Hekker} S.,  2010, \mn@doi [\mnras]
  {10.1111/j.1365-2966.2010.17141.x}, \href
  {http://adsabs.harvard.edu/abs/2010MNRAS.408..542C} {408, 542}

\bibitem[\protect\citeauthoryear{{Campante} et~al.,}{{Campante}
  et~al.}{2015}]{Campante15}
{Campante} T.~L.,  et~al., 2015, \mn@doi [\apj] {10.1088/0004-637X/799/2/170},
  \href {http://adsabs.harvard.edu/abs/2015ApJ...799..170C} {799, 170}

\bibitem[\protect\citeauthoryear{{Campante} et~al.,}{{Campante}
  et~al.}{2016a}]{CampanteObliquities}
{Campante} T.~L.,  et~al., 2016a, \mn@doi [\apj] {10.3847/0004-637X/819/1/85},
  \href {http://adsabs.harvard.edu/abs/2016ApJ...819...85C} {819, 85}

\bibitem[\protect\citeauthoryear{{Campante} et~al.,}{{Campante}
  et~al.}{2016b}]{CampanteTESS}
{Campante} T.~L.,  et~al., 2016b, \mn@doi [\apj] {10.3847/0004-637X/830/2/138},
  \href {http://adsabs.harvard.edu/abs/2016ApJ...830..138C} {830, 138}

\bibitem[\protect\citeauthoryear{{Chaplin} \& {Miglio}}{{Chaplin} \&
  {Miglio}}{2013}]{ChaplinMiglio}
{Chaplin} W.~J.,  {Miglio} A.,  2013, \mn@doi [\araa]
  {10.1146/annurev-astro-082812-140938}, \href
  {http://adsabs.harvard.edu/abs/2013ARA%26A..51..353C} {51, 353}

\bibitem[\protect\citeauthoryear{{Chaplin} et~al.,}{{Chaplin}
  et~al.}{2011}]{ChaplinSci}
{Chaplin} W.~J.,  et~al., 2011, \mn@doi [Science] {10.1126/science.1201827},
  \href {http://adsabs.harvard.edu/abs/2011Sci...332..213C} {332, 213}

\bibitem[\protect\citeauthoryear{{Chaplin} et~al.,}{{Chaplin}
  et~al.}{2015}]{K2astero1}
{Chaplin} W.~J.,  et~al., 2015, \mn@doi [\pasp] {10.1086/683103}, \href
  {http://adsabs.harvard.edu/abs/2015PASP..127.1038C} {127, 1038}

\bibitem[\protect\citeauthoryear{{Creevey} et~al.,}{{Creevey}
  et~al.}{2012}]{Creevey12}
{Creevey} O.~L.,  et~al., 2012, \mn@doi [\aap] {10.1051/0004-6361/201117037},
  \href {http://adsabs.harvard.edu/abs/2012A%26A...537A.111C} {537, A111}

\bibitem[\protect\citeauthoryear{{Crepp} \& {Johnson}}{{Crepp} \&
  {Johnson}}{2011}]{CreppJohnson}
{Crepp} J.~R.,  {Johnson} J.~A.,  2011, \mn@doi [\apj]
  {10.1088/0004-637X/733/2/126}, \href
  {http://adsabs.harvard.edu/abs/2011ApJ...733..126C} {733, 126}

\bibitem[\protect\citeauthoryear{{Davies} \& {Miglio}}{{Davies} \&
  {Miglio}}{2016}]{DaviesMiglio}
{Davies} G.~R.,  {Miglio} A.,  2016, \mn@doi [Astronomische Nachrichten]
  {10.1002/asna.201612371}, \href
  {http://adsabs.harvard.edu/abs/2016AN....337..774D} {337, 774}

\bibitem[\protect\citeauthoryear{{Davies} et~al.,}{{Davies}
  et~al.}{2016}]{Davies16}
{Davies} G.~R.,  et~al., 2016, \mn@doi [\mnras] {10.1093/mnras/stv2593}, \href
  {http://adsabs.harvard.edu/abs/2016MNRAS.456.2183D} {456, 2183}

\bibitem[\protect\citeauthoryear{{Demarque}, {Woo}, {Kim}  \& {Yi}}{{Demarque}
  et~al.}{2004}]{Y2}
{Demarque} P.,  {Woo} J.-H.,  {Kim} Y.-C.,   {Yi} S.~K.,  2004, \mn@doi [\apjs]
  {10.1086/424966}, \href {http://adsabs.harvard.edu/abs/2004ApJS..155..667D}
  {155, 667}

\bibitem[\protect\citeauthoryear{{Ferguson}, {Alexander}, {Allard}, {Barman},
  {Bodnarik}, {Hauschildt}, {Heffner-Wong}  \& {Tamanai}}{{Ferguson}
  et~al.}{2005}]{Ferguson05}
{Ferguson} J.~W.,  {Alexander} D.~R.,  {Allard} F.,  {Barman} T.,  {Bodnarik}
  J.~G.,  {Hauschildt} P.~H.,  {Heffner-Wong} A.,   {Tamanai} A.,  2005,
  \mn@doi [\apj] {10.1086/428642}, \href
  {http://adsabs.harvard.edu/abs/2005ApJ...623..585F} {623, 585}

\bibitem[\protect\citeauthoryear{{Flower}}{{Flower}}{1996}]{Flower96}
{Flower} P.~J.,  1996, \mn@doi [\apj] {10.1086/177785}, \href
  {http://adsabs.harvard.edu/abs/1996ApJ...469..355F} {469, 355}

\bibitem[\protect\citeauthoryear{{Gaia Collaboration} et~al.,}{{Gaia
  Collaboration} et~al.}{2016}]{BrownGaia}
{Gaia Collaboration} et~al., 2016, \mn@doi [\aap]
  {10.1051/0004-6361/201629512}, \href
  {http://adsabs.harvard.edu/abs/2016A%26A...595A...2G} {595, A2}

\bibitem[\protect\citeauthoryear{{Garc{\'{\i}}a} et~al.,}{{Garc{\'{\i}}a}
  et~al.}{2014}]{Garcia14}
{Garc{\'{\i}}a} R.~A.,  et~al., 2014, \mn@doi [\aap]
  {10.1051/0004-6361/201423888}, \href
  {http://adsabs.harvard.edu/abs/2014A%26A...572A..34G} {572, A34}

\bibitem[\protect\citeauthoryear{{Gaulme} et~al.,}{{Gaulme}
  et~al.}{2016}]{Gaulme16}
{Gaulme} P.,  et~al., 2016, \mn@doi [\apj] {10.3847/0004-637X/832/2/121}, \href
  {http://adsabs.harvard.edu/abs/2016ApJ...832..121G} {832, 121}

\bibitem[\protect\citeauthoryear{{Ghezzi} \& {Johnson}}{{Ghezzi} \&
  {Johnson}}{2015}]{GhezziJohnson}
{Ghezzi} L.,  {Johnson} J.~A.,  2015, \mn@doi [\apj]
  {10.1088/0004-637X/812/2/96}, \href
  {http://adsabs.harvard.edu/abs/2015ApJ...812...96G} {812, 96}

\bibitem[\protect\citeauthoryear{{Green} et~al.,}{{Green}
  et~al.}{2015}]{Green15}
{Green} G.~M.,  et~al., 2015, \mn@doi [\apj] {10.1088/0004-637X/810/1/25},
  \href {http://adsabs.harvard.edu/abs/2015ApJ...810...25G} {810, 25}

\bibitem[\protect\citeauthoryear{{Grevesse} \& {Noels}}{{Grevesse} \&
  {Noels}}{1993}]{GN93}
{Grevesse} N.,  {Noels} A.,  1993, \mn@doi [Physica Scripta Volume T]
  {10.1088/0031-8949/1993/T47/021}, \href
  {http://adsabs.harvard.edu/abs/1993PhST...47..133G} {47, 133}

\bibitem[\protect\citeauthoryear{{Grunblatt} et~al.,}{{Grunblatt}
  et~al.}{2016}]{Grunblatt}
{Grunblatt} S.~K.,  et~al., 2016, \mn@doi [\aj] {10.3847/0004-6256/152/6/185},
  \href {http://adsabs.harvard.edu/abs/2016AJ....152..185G} {152, 185}

\bibitem[\protect\citeauthoryear{{Handberg} \& {Lund}}{{Handberg} \&
  {Lund}}{2014}]{HandLund}
{Handberg} R.,  {Lund} M.~N.,  2014, \mn@doi [\mnras] {10.1093/mnras/stu1823},
  \href {http://adsabs.harvard.edu/abs/2014MNRAS.445.2698H} {445, 2698}

\bibitem[\protect\citeauthoryear{{Hatzes}, {Cochran}, {Endl}, {McArthur},
  {Paulson}, {Walker}, {Campbell}  \& {Yang}}{{Hatzes} et~al.}{2003}]{Hatzes03}
{Hatzes} A.~P.,  {Cochran} W.~D.,  {Endl} M.,  {McArthur} B.,  {Paulson} D.~B.,
   {Walker} G.~A.~H.,  {Campbell} B.,   {Yang} S.,  2003, \mn@doi [\apj]
  {10.1086/379281}, \href {http://adsabs.harvard.edu/abs/2003ApJ...599.1383H}
  {599, 1383}

\bibitem[\protect\citeauthoryear{{Hekker} et~al.,}{{Hekker}
  et~al.}{2010}]{Hekker10}
{Hekker} S.,  et~al., 2010, \mn@doi [\mnras]
  {10.1111/j.1365-2966.2009.16030.x}, \href
  {http://adsabs.harvard.edu/abs/2010MNRAS.402.2049H} {402, 2049}

\bibitem[\protect\citeauthoryear{{Hekker} et~al.,}{{Hekker}
  et~al.}{2011}]{Hekker11}
{Hekker} S.,  et~al., 2011, \mn@doi [\aap] {10.1051/0004-6361/201015185}, \href
  {http://adsabs.harvard.edu/abs/2011A%26A...525A.131H} {525, A131}

\bibitem[\protect\citeauthoryear{{Hj{\o}rringgaard}, {Silva Aguirre}, {White},
  {Huber}, {Pope}, {Casagrande}, {Justesen}  \&
  {Christensen-Dalsgaard}}{{Hj{\o}rringgaard} et~al.}{2016}]{retiredKpl}
{Hj{\o}rringgaard} J.~G.,  {Silva Aguirre} V.,  {White} T.~R.,  {Huber} D.,
  {Pope} B.~J.~S.,  {Casagrande} L.,  {Justesen} A.~B.,
  {Christensen-Dalsgaard} J.,  2016, \mn@doi [\mnras] {10.1093/mnras/stw2559},
  \href {http://adsabs.harvard.edu/abs/2016MNRAS.tmp.1542H} {}

\bibitem[\protect\citeauthoryear{{H{\o}g} et~al.,}{{H{\o}g}
  et~al.}{2000}]{Tycho2}
{H{\o}g} E.,  et~al., 2000, \aap, \href
  {http://cdsads.u-strasbg.fr/abs/2000A%26A...355L..27H} {355, L27}

\bibitem[\protect\citeauthoryear{{Houk} \& {Smith-Moore}}{{Houk} \&
  {Smith-Moore}}{1988}]{spectype}
{Houk} N.,  {Smith-Moore} M.,  1988, {Michigan Catalogue of Two-dimensional
  Spectral Types for the HD Stars. Volume 4, Declinations $-26{\fdg}0$ to
  $-12{\fdg}0$}.
Department of Astronomy, University of Michigan, USA

\bibitem[\protect\citeauthoryear{{Huber}, {Stello}, {Bedding}, {Chaplin},
  {Arentoft}, {Quirion}  \& {Kjeldsen}}{{Huber} et~al.}{2009}]{Huber09}
{Huber} D.,  {Stello} D.,  {Bedding} T.~R.,  {Chaplin} W.~J.,  {Arentoft} T.,
  {Quirion} P.-O.,   {Kjeldsen} H.,  2009, Communications in Asteroseismology,
  \href {http://adsabs.harvard.edu/abs/2009CoAst.160...74H} {160, 74}

\bibitem[\protect\citeauthoryear{{Huber} et~al.,}{{Huber}
  et~al.}{2013}]{Huber13}
{Huber} D.,  et~al., 2013, \mn@doi [\apj] {10.1088/0004-637X/767/2/127}, \href
  {http://adsabs.harvard.edu/abs/2013ApJ...767..127H} {767, 127}

\bibitem[\protect\citeauthoryear{{Huber} et~al.,}{{Huber} et~al.}{2016}]{EPIC}
{Huber} D.,  et~al., 2016, \mn@doi [\apjs] {10.3847/0067-0049/224/1/2}, \href
  {http://adsabs.harvard.edu/abs/2016ApJS..224....2H} {224, 2}

\bibitem[\protect\citeauthoryear{{Hurley}, {Pols}  \& {Tout}}{{Hurley}
  et~al.}{2000}]{huretal2000}
{Hurley} J.~R.,  {Pols} O.~R.,   {Tout} C.~A.,  2000, \mn@doi [\mnras]
  {10.1046/j.1365-8711.2000.03426.x}, \href
  {http://adsabs.harvard.edu/abs/2000MNRAS.315..543H} {315, 543}

\bibitem[\protect\citeauthoryear{{Iglesias} \& {Rogers}}{{Iglesias} \&
  {Rogers}}{1996}]{IR96}
{Iglesias} C.~A.,  {Rogers} F.~J.,  1996, \mn@doi [\apj] {10.1086/177381},
  \href {http://adsabs.harvard.edu/abs/1996ApJ...464..943I} {464, 943}

\bibitem[\protect\citeauthoryear{{Jofr{\'e}}, {Petrucci}, {Saffe}, {Saker}, {de
  la Villarmois}, {Chavero}, {G{\'o}mez}  \& {Mauas}}{{Jofr{\'e}}
  et~al.}{2015}]{Jofre15}
{Jofr{\'e}} E.,  {Petrucci} R.,  {Saffe} C.,  {Saker} L.,  {de la Villarmois}
  E.~A.,  {Chavero} C.,  {G{\'o}mez} M.,   {Mauas} P.~J.~D.,  2015, \mn@doi
  [\aap] {10.1051/0004-6361/201424474}, \href
  {http://adsabs.harvard.edu/abs/2015A%26A...574A..50J} {574, A50}

\bibitem[\protect\citeauthoryear{{Johnson} et~al.,}{{Johnson}
  et~al.}{2007a}]{Johnson07}
{Johnson} J.~A.,  et~al., 2007a, \mn@doi [\apj] {10.1086/519677}, \href
  {http://adsabs.harvard.edu/abs/2007ApJ...665..785J} {665, 785}

\bibitem[\protect\citeauthoryear{{Johnson}, {Butler}, {Marcy}, {Fischer},
  {Vogt}, {Wright}  \& {Peek}}{{Johnson} et~al.}{2007b}]{JohnsonMdwarf}
{Johnson} J.~A.,  {Butler} R.~P.,  {Marcy} G.~W.,  {Fischer} D.~A.,  {Vogt}
  S.~S.,  {Wright} J.~T.,   {Peek} K.~M.~G.,  2007b, \mn@doi [\apj]
  {10.1086/521720}, \href {http://adsabs.harvard.edu/abs/2007ApJ...670..833J}
  {670, 833}

\bibitem[\protect\citeauthoryear{{Johnson}, {Howard}, {Bowler}, {Henry},
  {Marcy}, {Wright}, {Fischer}  \& {Isaacson}}{{Johnson}
  et~al.}{2010a}]{discovery}
{Johnson} J.~A.,  {Howard} A.~W.,  {Bowler} B.~P.,  {Henry} G.~W.,  {Marcy}
  G.~W.,  {Wright} J.~T.,  {Fischer} D.~A.,   {Isaacson} H.,  2010a, \mn@doi
  [\pasp] {10.1086/653809}, \href
  {http://adsabs.harvard.edu/abs/2010PASP..122..701J} {122, 701}

\bibitem[\protect\citeauthoryear{{Johnson}, {Aller}, {Howard}  \&
  {Crepp}}{{Johnson} et~al.}{2010b}]{Johnson10}
{Johnson} J.~A.,  {Aller} K.~M.,  {Howard} A.~W.,   {Crepp} J.~R.,  2010b,
  \mn@doi [\pasp] {10.1086/655775}, \href
  {http://adsabs.harvard.edu/abs/2010PASP..122..905J} {122, 905}

\bibitem[\protect\citeauthoryear{{Johnson}, {Morton}  \& {Wright}}{{Johnson}
  et~al.}{2013}]{Johnson13}
{Johnson} J.~A.,  {Morton} T.~D.,   {Wright} J.~T.,  2013, \mn@doi [\apj]
  {10.1088/0004-637X/763/1/53}, \href
  {http://adsabs.harvard.edu/abs/2013ApJ...763...53J} {763, 53}

\bibitem[\protect\citeauthoryear{{Johnson} et~al.,}{{Johnson}
  et~al.}{2014}]{Johnson14}
{Johnson} J.~A.,  et~al., 2014, \mn@doi [\apj] {10.1088/0004-637X/794/1/15},
  \href {http://adsabs.harvard.edu/abs/2014ApJ...794...15J} {794, 15}

\bibitem[\protect\citeauthoryear{{Kaufer}, {Stahl}, {Tubbesing},
  {N{\o}rregaard}, {Avila}, {Francois}, {Pasquini}  \& {Pizzella}}{{Kaufer}
  et~al.}{1999}]{FEROS}
{Kaufer} A.,  {Stahl} O.,  {Tubbesing} S.,  {N{\o}rregaard} P.,  {Avila} G.,
  {Francois} P.,  {Pasquini} L.,   {Pizzella} A.,  1999, The Messenger, \href
  {http://adsabs.harvard.edu/abs/1999Msngr..95....8K} {95, 8}

\bibitem[\protect\citeauthoryear{{Kjeldsen} \& {Bedding}}{{Kjeldsen} \&
  {Bedding}}{1995}]{KB95}
{Kjeldsen} H.,  {Bedding} T.~R.,  1995, \aap, \href
  {http://adsabs.harvard.edu/abs/1995A%26A...293...87K} {293, 87}

\bibitem[\protect\citeauthoryear{{Krishna Swamy}}{{Krishna Swamy}}{1966}]{KS66}
{Krishna Swamy} K.~S.,  1966, \mn@doi [\apj] {10.1086/148752}, \href
  {http://adsabs.harvard.edu/abs/1966ApJ...145..174K} {145, 174}

\bibitem[\protect\citeauthoryear{{Lagarde}, {Decressin}, {Charbonnel},
  {Eggenberger}, {Ekstr{\"o}m}  \& {Palacios}}{{Lagarde}
  et~al.}{2012}]{Lagarde12}
{Lagarde} N.,  {Decressin} T.,  {Charbonnel} C.,  {Eggenberger} P.,
  {Ekstr{\"o}m} S.,   {Palacios} A.,  2012, \mn@doi [\aap]
  {10.1051/0004-6361/201118331}, \href
  {http://adsabs.harvard.edu/abs/2012A%26A...543A.108L} {543, A108}

\bibitem[\protect\citeauthoryear{{Lagarde}, {Bossini}, {Miglio}, {Vrard}  \&
  {Mosser}}{{Lagarde} et~al.}{2016}]{Lagarde16}
{Lagarde} N.,  {Bossini} D.,  {Miglio} A.,  {Vrard} M.,   {Mosser} B.,  2016,
  \mn@doi [\mnras] {10.1093/mnrasl/slv201}, \href
  {http://adsabs.harvard.edu/abs/2016MNRAS.457L..59L} {457, L59}

\bibitem[\protect\citeauthoryear{{Lind}, {Asplund}  \& {Barklem}}{{Lind}
  et~al.}{2009}]{lind09}
{Lind} K.,  {Asplund} M.,   {Barklem} P.~S.,  2009, \mn@doi [\aap]
  {10.1051/0004-6361/200912221}, \href
  {http://adsabs.harvard.edu/abs/2009A%26A...503..541L} {503, 541}

\bibitem[\protect\citeauthoryear{{Lindegren} et~al.,}{{Lindegren}
  et~al.}{2016}]{GaiaDR1}
{Lindegren} L.,  et~al., 2016, \mn@doi [\aap] {10.1051/0004-6361/201628714},
  \href {http://adsabs.harvard.edu/abs/2016A%26A...595A...4L} {595, A4}

\bibitem[\protect\citeauthoryear{{Lloyd}}{{Lloyd}}{2011}]{Lloyd11}
{Lloyd} J.~P.,  2011, \mn@doi [\apjl] {10.1088/2041-8205/739/2/L49}, \href
  {http://adsabs.harvard.edu/abs/2011ApJ...739L..49L} {739, L49}

\bibitem[\protect\citeauthoryear{{Lloyd}}{{Lloyd}}{2013}]{Lloyd13}
{Lloyd} J.~P.,  2013, \mn@doi [\apjl] {10.1088/2041-8205/774/1/L2}, \href
  {http://adsabs.harvard.edu/abs/2013ApJ...774L...2L} {774, L2}

\bibitem[\protect\citeauthoryear{{Lovis} \& {Mayor}}{{Lovis} \&
  {Mayor}}{2007}]{LovisMayor}
{Lovis} C.,  {Mayor} M.,  2007, \mn@doi [\aap] {10.1051/0004-6361:20077375},
  \href {http://adsabs.harvard.edu/abs/2007A%26A...472..657L} {472, 657}

\bibitem[\protect\citeauthoryear{{Lund}, {Handberg}, {Davies}, {Chaplin}  \&
  {Jones}}{{Lund} et~al.}{2015}]{K2P2}
{Lund} M.~N.,  {Handberg} R.,  {Davies} G.~R.,  {Chaplin} W.~J.,   {Jones}
  C.~D.,  2015, \mn@doi [\apj] {10.1088/0004-637X/806/1/30}, \href
  {http://adsabs.harvard.edu/abs/2015ApJ...806...30L} {806, 30}

\bibitem[\protect\citeauthoryear{{Lund} et~al.,}{{Lund} et~al.}{2016a}]{Lund16}
{Lund} M.~N.,  et~al., 2016a, \mn@doi [\pasp]
  {10.1088/1538-3873/128/970/124204}, \href
  {http://adsabs.harvard.edu/abs/2016PASP..128l4204L} {128, 124204}

\bibitem[\protect\citeauthoryear{{Lund} et~al.,}{{Lund}
  et~al.}{2016b}]{LundHyades}
{Lund} M.~N.,  et~al., 2016b, \mn@doi [\mnras] {10.1093/mnras/stw2160}, \href
  {http://adsabs.harvard.edu/abs/2016MNRAS.463.2600L} {463, 2600}

\bibitem[\protect\citeauthoryear{{Lundkvist} et~al.,}{{Lundkvist}
  et~al.}{2016}]{Lundkvist}
{Lundkvist} M.~S.,  et~al., 2016, \mn@doi [Nature Communications]
  {10.1038/ncomms11201}, \href
  {http://adsabs.harvard.edu/abs/2016NatCo...711201L} {7, 11201}

\bibitem[\protect\citeauthoryear{{Maeder}}{{Maeder}}{1975}]{Maeder75}
{Maeder} A.,  1975, \aap, \href
  {http://adsabs.harvard.edu/abs/1975A%26A....40..303M} {40, 303}

\bibitem[\protect\citeauthoryear{{Maldonado}, {Villaver}  \&
  {Eiroa}}{{Maldonado} et~al.}{2013}]{Maldonado13}
{Maldonado} J.,  {Villaver} E.,   {Eiroa} C.,  2013, \mn@doi [\aap]
  {10.1051/0004-6361/201321082}, \href
  {http://adsabs.harvard.edu/abs/2013A%26A...554A..84M} {554, A84}

\bibitem[\protect\citeauthoryear{{Masseron} \& {Hawkins}}{{Masseron} \&
  {Hawkins}}{2017}]{Masseron}
{Masseron} T.,  {Hawkins} K.,  2017, \mn@doi [\aap]
  {10.1051/0004-6361/201629938}, \href
  {http://adsabs.harvard.edu/abs/2017A%26A...597L...3M} {597, L3}

\bibitem[\protect\citeauthoryear{{Michalik}, {Lindegren}  \&
  {Hobbs}}{{Michalik} et~al.}{2015}]{Michalik15}
{Michalik} D.,  {Lindegren} L.,   {Hobbs} D.,  2015, \mn@doi [\aap]
  {10.1051/0004-6361/201425310}, \href
  {http://adsabs.harvard.edu/abs/2015A%26A...574A.115M} {574, A115}

\bibitem[\protect\citeauthoryear{{Miglio} et~al.,}{{Miglio}
  et~al.}{2016}]{Miglio16}
{Miglio} A.,  et~al., 2016, \mn@doi [\mnras] {10.1093/mnras/stw1555}, \href
  {http://adsabs.harvard.edu/abs/2016MNRAS.461..760M} {461, 760}

\bibitem[\protect\citeauthoryear{{Morel} et~al.,}{{Morel}
  et~al.}{2014}]{morel14}
{Morel} T.,  et~al., 2014, \mn@doi [\aap] {10.1051/0004-6361/201322810}, \href
  {http://adsabs.harvard.edu/abs/2014A%26A...564A.119M} {564, A119}

\bibitem[\protect\citeauthoryear{{Mortier}, {Santos}, {Sousa}, {Adibekyan},
  {Delgado Mena}, {Tsantaki}, {Israelian}  \& {Mayor}}{{Mortier}
  et~al.}{2013}]{Mortier13}
{Mortier} A.,  {Santos} N.~C.,  {Sousa} S.~G.,  {Adibekyan} V.~Z.,  {Delgado
  Mena} E.,  {Tsantaki} M.,  {Israelian} G.,   {Mayor} M.,  2013, \mn@doi
  [\aap] {10.1051/0004-6361/201321641}, \href
  {http://adsabs.harvard.edu/abs/2013A%26A...557A..70M} {557, A70}

\bibitem[\protect\citeauthoryear{{Mosser} et~al.,}{{Mosser}
  et~al.}{2011}]{Mosser11}
{Mosser} B.,  et~al., 2011, \mn@doi [\aap] {10.1051/0004-6361/201015440}, \href
  {http://adsabs.harvard.edu/abs/2011A%26A...525L...9M} {525, L9}

\bibitem[\protect\citeauthoryear{{Mosser} et~al.,}{{Mosser}
  et~al.}{2012}]{Mosser12}
{Mosser} B.,  et~al., 2012, \mn@doi [\aap] {10.1051/0004-6361/201118519}, \href
  {http://adsabs.harvard.edu/abs/2012A%26A...540A.143M} {540, A143}

\bibitem[\protect\citeauthoryear{{Mustill} \& {Villaver}}{{Mustill} \&
  {Villaver}}{2012}]{musvil2012}
{Mustill} A.~J.,  {Villaver} E.,  2012, \mn@doi [\apj]
  {10.1088/0004-637X/761/2/121}, \href
  {http://ukads.nottingham.ac.uk/abs/2012ApJ...761..121M} {761, 121}

\bibitem[\protect\citeauthoryear{{Niedzielski} et~al.,}{{Niedzielski}
  et~al.}{2015}]{TAPAS}
{Niedzielski} A.,  et~al., 2015, \mn@doi [\aap] {10.1051/0004-6361/201424399},
  \href {http://adsabs.harvard.edu/abs/2015A%26A...573A..36N} {573, A36}

\bibitem[\protect\citeauthoryear{{Nordhaus} \& {Spiegel}}{{Nordhaus} \&
  {Spiegel}}{2013}]{norspi2013}
{Nordhaus} J.,  {Spiegel} D.~S.,  2013, \mn@doi [\mnras]
  {10.1093/mnras/stt569}, \href
  {http://ukads.nottingham.ac.uk/abs/2013MNRAS.432..500N} {432, 500}

\bibitem[\protect\citeauthoryear{{Osaki}}{{Osaki}}{1975}]{Osaki75}
{Osaki} J.,  1975, \pasj, \href
  {http://adsabs.harvard.edu/abs/1975PASJ...27..237O} {27, 237}

\bibitem[\protect\citeauthoryear{{Paxton}, {Bildsten}, {Dotter}, {Herwig},
  {Lesaffre}  \& {Timmes}}{{Paxton} et~al.}{2011}]{Paxton11}
{Paxton} B.,  {Bildsten} L.,  {Dotter} A.,  {Herwig} F.,  {Lesaffre} P.,
  {Timmes} F.,  2011, \mn@doi [\apjs] {10.1088/0067-0049/192/1/3}, \href
  {http://adsabs.harvard.edu/abs/2011ApJS..192....3P} {192, 3}

\bibitem[\protect\citeauthoryear{{Paxton} et~al.,}{{Paxton}
  et~al.}{2013}]{Paxton13}
{Paxton} B.,  et~al., 2013, \mn@doi [\apjs] {10.1088/0067-0049/208/1/4}, \href
  {http://adsabs.harvard.edu/abs/2013ApJS..208....4P} {208, 4}

\bibitem[\protect\citeauthoryear{{Pijpers}}{{Pijpers}}{2003}]{Pijpers03}
{Pijpers} F.~P.,  2003, \mn@doi [\aap] {10.1051/0004-6361:20021839}, \href
  {http://adsabs.harvard.edu/abs/2003A%26A...400..241P} {400, 241}

\bibitem[\protect\citeauthoryear{{Pinsonneault} et~al.,}{{Pinsonneault}
  et~al.}{2014}]{Pinsonneault14}
{Pinsonneault} M.~H.,  et~al., 2014, \mn@doi [\apjs]
  {10.1088/0067-0049/215/2/19}, \href
  {http://adsabs.harvard.edu/abs/2014ApJS..215...19P} {215, 19}

\bibitem[\protect\citeauthoryear{{Reffert}, {Quirrenbach}, {Mitchell},
  {Albrecht}, {Hekker}, {Fischer}, {Marcy}  \& {Butler}}{{Reffert}
  et~al.}{2006}]{Reffert06}
{Reffert} S.,  {Quirrenbach} A.,  {Mitchell} D.~S.,  {Albrecht} S.,  {Hekker}
  S.,  {Fischer} D.~A.,  {Marcy} G.~W.,   {Butler} R.~P.,  2006, \mn@doi [\apj]
  {10.1086/507516}, \href {http://adsabs.harvard.edu/abs/2006ApJ...652..661R}
  {652, 661}

\bibitem[\protect\citeauthoryear{{Ricker} et~al.,}{{Ricker}
  et~al.}{2015}]{TESS}
{Ricker} G.~R.,  et~al., 2015, \mn@doi [Journal of Astronomical Telescopes,
  Instruments, and Systems] {10.1117/1.JATIS.1.1.014003}, \href
  {http://adsabs.harvard.edu/abs/2015JATIS...1a4003R} {1, 014003}

\bibitem[\protect\citeauthoryear{{Rodrigues} et~al.,}{{Rodrigues}
  et~al.}{2014}]{PARAM2}
{Rodrigues} T.~S.,  et~al., 2014, \mn@doi [\mnras] {10.1093/mnras/stu1907},
  \href {http://adsabs.harvard.edu/abs/2014MNRAS.445.2758R} {445, 2758}

\bibitem[\protect\citeauthoryear{{Rodrigues} et~al.,}{{Rodrigues}
  et~al.}{2017}]{Rodrigues17}
{Rodrigues} T.~S.,  et~al., 2017, \mn@doi [\mnras] {10.1093/mnras/stx120},
  \href {http://adsabs.harvard.edu/abs/2017MNRAS.tmp..120R} {467, 1433}

\bibitem[\protect\citeauthoryear{{Rogers} \& {Nayfonov}}{{Rogers} \&
  {Nayfonov}}{2002}]{RN02}
{Rogers} F.~J.,  {Nayfonov} A.,  2002, \mn@doi [\apj] {10.1086/341894}, \href
  {http://adsabs.harvard.edu/abs/2002ApJ...576.1064R} {576, 1064}

\bibitem[\protect\citeauthoryear{{Sackmann}, {Boothroyd}  \&
  {Kraemer}}{{Sackmann} et~al.}{1993}]{sacetal1993}
{Sackmann} I.-J.,  {Boothroyd} A.~I.,   {Kraemer} K.~E.,  1993, \mn@doi [\apj]
  {10.1086/173407}, \href {http://adsabs.harvard.edu/abs/1993ApJ...418..457S}
  {418, 457}

\bibitem[\protect\citeauthoryear{{Santos} et~al.,}{{Santos}
  et~al.}{2013}]{Santos13}
{Santos} N.~C.,  et~al., 2013, \mn@doi [\aap] {10.1051/0004-6361/201321286},
  \href {http://adsabs.harvard.edu/abs/2013A%26A...556A.150S} {556, A150}

\bibitem[\protect\citeauthoryear{{Sato} et~al.,}{{Sato} et~al.}{2008}]{Sato08}
{Sato} B.,  et~al., 2008, \mn@doi [\pasj] {10.1093/pasj/60.3.539}, \href
  {http://adsabs.harvard.edu/abs/2008PASJ...60..539S} {60, 539}

\bibitem[\protect\citeauthoryear{{Schlaufman} \& {Winn}}{{Schlaufman} \&
  {Winn}}{2013}]{Schlaufman13}
{Schlaufman} K.~C.,  {Winn} J.~N.,  2013, \mn@doi [\apj]
  {10.1088/0004-637X/772/2/143}, \href
  {http://adsabs.harvard.edu/abs/2013ApJ...772..143S} {772, 143}

\bibitem[\protect\citeauthoryear{{Schuler}, {Hatzes}, {King}, {K{\"u}rster}  \&
  {The}}{{Schuler} et~al.}{2006}]{schuler06}
{Schuler} S.~C.,  {Hatzes} A.~P.,  {King} J.~R.,  {K{\"u}rster} M.,   {The}
  L.-S.,  2006, \mn@doi [\aj] {10.1086/499103}, \href
  {http://adsabs.harvard.edu/abs/2006AJ....131.1057S} {131, 1057}

\bibitem[\protect\citeauthoryear{{Setiawan} et~al.,}{{Setiawan}
  et~al.}{2005}]{Setiawan05}
{Setiawan} J.,  et~al., 2005, \mn@doi [\aap] {10.1051/0004-6361:200500133},
  \href {http://adsabs.harvard.edu/abs/2005A%26A...437L..31S} {437, L31}

\bibitem[\protect\citeauthoryear{{Sharma}, {Stello}, {Bland-Hawthorn}, {Huber}
  \& {Bedding}}{{Sharma} et~al.}{2016}]{Sharma16}
{Sharma} S.,  {Stello} D.,  {Bland-Hawthorn} J.,  {Huber} D.,   {Bedding}
  T.~R.,  2016, \mn@doi [\apj] {10.3847/0004-637X/822/1/15}, \href
  {http://adsabs.harvard.edu/abs/2016ApJ...822...15S} {822, 15}

\bibitem[\protect\citeauthoryear{{Silva Aguirre} et~al.,}{{Silva Aguirre}
  et~al.}{2015}]{VSA15}
{Silva Aguirre} V.,  et~al., 2015, \mn@doi [\mnras] {10.1093/mnras/stv1388},
  \href {http://adsabs.harvard.edu/abs/2015MNRAS.452.2127S} {452, 2127}

\bibitem[\protect\citeauthoryear{{Staff}, {De Marco}, {Wood}, {Galaviz}  \&
  {Passy}}{{Staff} et~al.}{2016}]{staetal2016}
{Staff} J.~E.,  {De Marco} O.,  {Wood} P.,  {Galaviz} P.,   {Passy} J.-C.,
  2016, \mn@doi [\mnras] {10.1093/mnras/stw331}, \href
  {http://ukads.nottingham.ac.uk/abs/2016MNRAS.458..832S} {458, 832}

\bibitem[\protect\citeauthoryear{{Stello} et~al.,}{{Stello}
  et~al.}{2009}]{Stello09}
{Stello} D.,  et~al., 2009, \mn@doi [\apj] {10.1088/0004-637X/700/2/1589},
  \href {http://adsabs.harvard.edu/abs/2009ApJ...700.1589S} {700, 1589}

\bibitem[\protect\citeauthoryear{{Stello} et~al.,}{{Stello}
  et~al.}{2015}]{K2astero2}
{Stello} D.,  et~al., 2015, \mn@doi [\apjl] {10.1088/2041-8205/809/1/L3}, \href
  {http://adsabs.harvard.edu/abs/2015ApJ...809L...3S} {809, L3}

\bibitem[\protect\citeauthoryear{{Stello} et~al.,}{{Stello}
  et~al.}{2016}]{Stello16}
{Stello} D.,  et~al., 2016, \mn@doi [\apj] {10.3847/0004-637X/832/2/133}, \href
  {http://adsabs.harvard.edu/abs/2016ApJ...832..133S} {832, 133}

\bibitem[\protect\citeauthoryear{{Torres}}{{Torres}}{2010}]{Torres10}
{Torres} G.,  2010, \mn@doi [\aj] {10.1088/0004-6256/140/5/1158}, \href
  {http://adsabs.harvard.edu/abs/2010AJ....140.1158T} {140, 1158}

\bibitem[\protect\citeauthoryear{{Torres}, {Fischer}, {Sozzetti}, {Buchhave},
  {Winn}, {Holman}  \& {Carter}}{{Torres} et~al.}{2012}]{Torres12}
{Torres} G.,  {Fischer} D.~A.,  {Sozzetti} A.,  {Buchhave} L.~A.,  {Winn}
  J.~N.,  {Holman} M.~J.,   {Carter} J.~A.,  2012, \mn@doi [\apj]
  {10.1088/0004-637X/757/2/161}, \href
  {http://adsabs.harvard.edu/abs/2012ApJ...757..161T} {757, 161}

\bibitem[\protect\citeauthoryear{{Tremblay}, {Cummings}, {Kalirai},
  {G{\"a}nsicke}, {Gentile-Fusillo}  \& {Raddi}}{{Tremblay}
  et~al.}{2016}]{treetal2016}
{Tremblay} P.-E.,  {Cummings} J.,  {Kalirai} J.~S.,  {G{\"a}nsicke} B.~T.,
  {Gentile-Fusillo} N.,   {Raddi} R.,  2016, \mn@doi [\mnras]
  {10.1093/mnras/stw1447}, \href
  {http://adsabs.harvard.edu/abs/2016MNRAS.461.2100T} {461, 2100}

\bibitem[\protect\citeauthoryear{{Ulrich}}{{Ulrich}}{1986}]{Ulrich86}
{Ulrich} R.~K.,  1986, \mn@doi [\apjl] {10.1086/184700}, \href
  {http://adsabs.harvard.edu/abs/1986ApJ...306L..37U} {306, L37}

\bibitem[\protect\citeauthoryear{{Veras}}{{Veras}}{2016}]{veras2016a}
{Veras} D.,  2016, \mn@doi [Royal Society Open Science] {10.1098/rsos.150571},
  \href {http://adsabs.harvard.edu/abs/2016RSOS....350571V} {3, 150571}

\bibitem[\protect\citeauthoryear{{Veras}, {Wyatt}, {Mustill}, {Bonsor}  \&
  {Eldridge}}{{Veras} et~al.}{2011}]{veretal2011}
{Veras} D.,  {Wyatt} M.~C.,  {Mustill} A.~J.,  {Bonsor} A.,   {Eldridge} J.~J.,
   2011, \mn@doi [\mnras] {10.1111/j.1365-2966.2011.19393.x}, \href
  {http://adsabs.harvard.edu/abs/2011MNRAS.417.2104V} {417, 2104}

\bibitem[\protect\citeauthoryear{{Veras}, {Mustill}  \& {G{\"a}nsicke}}{{Veras}
  et~al.}{2017}]{veretal2016b}
{Veras} D.,  {Mustill} A.~J.,   {G{\"a}nsicke} B.~T.,  2017, \mn@doi [\mnras]
  {10.1093/mnras/stw2821}, \href
  {http://adsabs.harvard.edu/abs/2017MNRAS.465.1499V} {465, 1499}

\bibitem[\protect\citeauthoryear{{Wittenmyer}, {Endl}, {Wang}, {Johnson},
  {Tinney}  \& {O'Toole}}{{Wittenmyer} et~al.}{2011}]{Wittenmyer11}
{Wittenmyer} R.~A.,  {Endl} M.,  {Wang} L.,  {Johnson} J.~A.,  {Tinney} C.~G.,
   {O'Toole} S.~J.,  2011, \mn@doi [\apj] {10.1088/0004-637X/743/2/184}, \href
  {http://adsabs.harvard.edu/abs/2011ApJ...743..184W} {743, 184}

\bibitem[\protect\citeauthoryear{{da Silva} et~al.,}{{da Silva}
  et~al.}{2006}]{PARAM1}
{da Silva} L.,  et~al., 2006, \mn@doi [\aap] {10.1051/0004-6361:20065105},
  \href {http://adsabs.harvard.edu/abs/2006A%26A...458..609D} {458, 609}

\bibitem[\protect\citeauthoryear{{do Nascimento}, {da Costa}  \& {Castro}}{{do
  Nascimento} et~al.}{2012}]{doNascimento12}
{do Nascimento} J.-D.,  {da Costa} J.~S.,   {Castro} M.,  2012, \mn@doi [\aap]
  {10.1051/0004-6361/201219791}, \href
  {http://adsabs.harvard.edu/abs/2012A%26A...548L...1D} {548, L1}

\bibitem[\protect\citeauthoryear{{van Leeuwen}}{{van
  Leeuwen}}{2007}]{Hipparcos}
{van Leeuwen} F.,  2007, \mn@doi [\aap] {10.1051/0004-6361:20078357}, \href
  {http://cdsads.u-strasbg.fr/abs/2007A%26A...474..653V} {474, 653}

\makeatother
\end{thebibliography}






\bsp	
\label{lastpage}
\end{document}